\documentclass[fleqn,usenatbib]{mnras}
\usepackage[T1]{fontenc}
\usepackage{newtxtext,newtxmath}
\usepackage{graphicx}
\usepackage{amsmath}
\usepackage{booktabs}
\usepackage{multirow}
\usepackage[normalem]{ulem}
\usepackage{xcolor}
\usepackage{tikz}
\usepackage{pgfplots}
\usepackage{url}
\pgfplotsset{compat=1.18}

\newcommand{\HI}{H{\small\,I}}
\newcommand{\kms}{\,\mathrm{km\,s^{-1}}}
\newcommand{\ms}{\,\mathrm{m\,s^{-2}}}
\definecolor{obsred}{HTML}{F8766D}
\definecolor{newtgreen}{HTML}{7CAE00}
\definecolor{mondcyan}{HTML}{00BFC4}
\definecolor{hmgpurple}{HTML}{C77CFF}
\definecolor{limitred}{RGB}{214,39,40}

\title[Gas-rich UDGs: alleviating the MOND tension with HMG]{Gas-rich ultra-diffuse galaxies: alleviating the MOND tension with hyperconical modified gravity}

\author[R. Monjo]{
Robert Monjo$^{1,2}$\thanks{E-mail: robert.monjo@cunef.edu}
\\
$^{1}$Department of Quantitative Methods, CUNEF University, Calle Almansa 101, 28040 Madrid, Spain\\
$^{2}$Centro de Estudios Cosmol\'ogicos ``Prof. Jaime Roessler Bonzi'', Facultad de Ciencias, Universidad de Chile, Santiago, Chile
}

\date{Accepted XXX. Received YYY; in original form ZZZ}
\pubyear{2026}

\begin{document}
\maketitle
\begin{abstract}
Gas-rich ultra-diffuse galaxies (UDGs) provide an unusually sharp test of gravity models tied to the baryonic Tully--Fisher relation because several systems appear to rotate too slowly for their baryonic masses. This study revisits the six isolated gas-rich UDGs analysed by  Mancera Piña \textit{et al}. with the current outer-radius prescription of hyperconical modified gravity (HMG), using the published baryonic masses, outer circular velocities, and Newtonian disc-based comparison velocities. The scan over the neighbourhood-scale parameter drives the model towards the asymptotic branch of HMG. For that limit, the HMG velocities are still systematically high for four of the six galaxies. Relative to the published central values and asymmetric $1\sigma$ velocity errors, the fixed asymptotic branch gives $\chi^2 \simeq 17.1$ for six objects, whereas the published Newtonian outer-radius disc estimates give $\chi^2 \simeq 9.3$ and a standard MOND interpolation is much worse ($\chi^2 \simeq 615.7$). Using combined observational and model uncertainties, the per-galaxy HMG tension ranges from $0.2\sigma$ to $2.0\sigma$, very similar to the $0.1\sigma$ to $1.7\sigma$ found for the Newtonian disc-based values, and much smaller than the $3.7\sigma$ to $5.9\sigma$ obtained for MOND. The likelihood scan showed that the preferred solutions lie on the asymptotic weak-coupling branch. We conclude that the present outer-radius HMG implementation alleviates the difficulties of MOND, but is still not sufficient to account for the published central values of the gas-rich UDG sample. Gas-rich UDGs therefore provide a useful discriminant between MOND and HMG.
\end{abstract}


\section{Introduction}

Gas-rich ultra-diffuse galaxies (UDGs) are an unusual low-surface-brightness population with large \HI\ reservoirs and extended stellar discs \citep{Leisman2017}. Their resolved kinematics suggested that several systems rotate too slowly for their baryonic masses, placing them below the canonical baryonic Tully--Fisher relation (BTFR) and leaving little room for dark matter within the observed radii \citep{ManceraPina2019,ManceraPina2020,ManceraPina2022}. Resolved Apertif data have also reinforced the view that UDG-like systems tend to occupy the low-velocity side of the dwarf-galaxy population \citep{Siljeg2024}. For instance, the updated 3D \HI\ modelling of AGC~114905 gives an outer circular speed of about $23^{+4}_{-6}\kms$, still low compared to the Newtonian expectation inferred from its baryons \citep{ManceraPina2022}. Such systems are therefore useful stress tests for both particle dark matter and modified-gravity scenarios. At the same time, recent work argued that the apparent discrepancy may be reduced once distance and inclination uncertainties are treated more conservatively \citep{Lelli2024}. A subsequent resolved discussion of AGC~114905 further emphasizes the role of geometry and distance assumptions \citep{ManceraPina2024}.

Modified Newtonian Dynamics (MOND) was the first successful attempt to model anomalous acceleration in late-type galaxies \citep{Milgrom1983,BanikZhao2022}, close to the cosmological \textit{de Sitter} scale $cH_0$ \citep{vanPutten2015,vanPutten2024}. This regime is especially relevant here because gas-rich UDGs probe low accelerations at large radii. Once the baryonic distribution and the interpolation function are specified, the low-acceleration outskirts of galaxies are expected to follow a tight MONDian scaling, so unusually low outer velocities become a sharp test.

In practice, it is convenient to write the true centripetal acceleration of MOND as $g_\text{\tiny MOND} =\hat{\nu}(g_\text{\tiny N }/a_0)\,g_\text{\tiny N }$, where $g_\text{\tiny N }$ is the Newtonian prediction from baryons and $\hat{\nu}$ is an inverse interpolation function. In the Newtonian regime, $\hat{\nu}\rightarrow1$ so that $g_\text{\tiny MOND} \rightarrow g_\text{\tiny N }$, whereas in the deep-MOND limit one obtains
\begin{equation}
    g_\text{\tiny MOND} \approx \sqrt{g_\text{\tiny N } a_0}\,.
\end{equation}
A standard selection of $\hat{\nu}$ was presented by McGaugh--Lelli--Schombert \citet{McGaugh2016}, derived from empirical fitting. This prescription is used below as the benchmark MOND comparison and is denoted hereafter as MLS. The MLS inverse function can be written as
\begin{equation}\label{eq:MLS}
\hat{\nu}_{\rm MLS}(x)=\left(1-e^{-\sqrt{x}}\right)^{-1},
\end{equation}
with $x=g_\text{\tiny N }/a_0$ and $a_0=1.2\times10^{-10}\ms$ \citep{BanikZhao2022}.

On the other hand, hyperconical modified gravity (HMG) is a relativistic MOND-like framework in which an apparent acceleration $a_{\gamma_0} = 2\gamma_0^{-1} cH_0$ emerges from the cosmological-system relative geometry \citep{Monjo2023CQG}, with projection angle $\gamma_0 = \gamma_s/\cos\gamma_s \sim 0.1$ that depends on a system parameter $\gamma_s \in [\pi/3,\, \pi/2]$. Unlike a universal MOND interpolation function, the effective HMG enhancement depends on this system angle and on the local-to-cosmic neighbourhood density contrast. This is the physical origin of the differences with respect to MOND that are most relevant for gas-rich UDGs.

The outer-radius prescription follows current formulations of the HMG model \citep{Monjo_2025_distinct,Monjo_2025hydro,Monjo_2025_weak}. The system angle $\gamma_s$ at the outer radius $r_{\rm sys}$ is written as
\begin{equation}\label{eq:gammaeps}
\sin^2\gamma_s(s) \approx \sin^2\gamma_U + \left(\sin^2\gamma_{\rm cen}-\sin^2\gamma_U\right)
\left|\frac{2v_\text{\tiny N }^2-\epsilon_s^2(s) v_\text{\tiny H}^2}{2v_\text{\tiny N }^2+\epsilon_s^2(s) v_\text{\tiny H}^2}\right|,
\end{equation}
where $\gamma_U=\pi/3$ and $\gamma_{\rm cen}=\pi/2$ denote, respectively, the minimum and maximum spatial limits of the hyperconical geometry, and $v_\text{\tiny H}=H_0 r_{\rm sys}$ is the Hubble expansion speed at $r_{\rm sys}$. The dimensionless quantity
\begin{equation}\label{eq:epss}
\epsilon_s^2(s) \equiv \frac{\rho_{{\rm nei}}(s)}{\rho_{\rm vac}} +\frac{1}{6}
= \frac{3 M_{\rm bar}}{4\pi s^3 r_{\rm sys}^3 \rho_{\rm vac}}+\frac{1}{6},
\end{equation}
measures the relative neighbourhood density $\rho_{{\rm nei}}(s)$ at the scale $s\,r_{\rm sys}$ with respect to the vacuum-density scale $\rho_{\rm vac}=3H_0^2/(8\pi G)$ and $s \ge 1$, while the term $1/6$ is the asymptotic constant of the HMG model \citep{Monjo_2025_distinct}. For gas-rich UDGs, the gas component can extend to radii of order ten times the optical disc scale length, making the asymptotic regime $\epsilon_s^2 \to 1/6$ relevant for the present analysis. The only scanned quantity is therefore the global neighbourhood-scale factor $s$, with solutions expected towards the limit $s \gg 1$.

In all cases, the total HMG centripetal acceleration can be written as
\begin{equation} \label{eq:HMG}
g_\text{\tiny HMG}(s) \approx \sqrt{g_\text{\tiny N }^2 + g_\text{\tiny N } a_{\gamma_0}(s)} =  \sqrt{g_\text{\tiny N }^2 + g_\text{\tiny N } {2c}H_0\frac{\cos\gamma_s(s)}{\gamma_s(s) }}.
\end{equation}

HMG has been tested in reproducing galaxy- and cluster-scale radial-acceleration relations with one free parameter. HMG can accommodate observations even from hydrostatic equilibrium of galaxy clusters in hot-gas-dominated regions and rotation curves inferred from weak lensing \citep{Monjo_2025hydro,Monjo_2025_weak}. These previous successes motivate testing whether the same framework can also describe gas-rich UDGs.

The present analysis therefore deliberately focuses on the outermost resolved radius $r_{\rm sys}$. This is not meant to replace full resolved-curve modelling. Rather, it provides a compact and homogeneous summary statistic across a small sample for which most galaxies have only a few independent kinematic points, while the innermost ones are more susceptible to beam smearing, inclination degeneracies, and non-circular motions. The outermost point is also the natural location where the gas-rich discs are most diffuse and where the asymptotic branch of the HMG outer-radius prescription should be most relevant. For AGC~114905, where better resolved data exist, a full rotation-curve fit remains a crucial next step and is discussed below.

The main aim of this paper is therefore to determine whether the current outer-radius HMG prescription can account for the published central values of the six UDGs analysed by \citet{ManceraPina2020}, and whether these galaxies discriminate between MOND and HMG in practice. The question is not yet whether HMG provides a definitive resolved description of every gas-rich UDG, but whether the current outer-radius prescription materially alleviates the strong MOND tension while remaining close to the published Newtonian disc-based estimates. That makes the present six-galaxy sample a compact, falsifiable, and easily updatable test as new kinematic data become available.

\section{Sample}

The sample consists of the six isolated \HI-bearing UDGs analysed by \citet{ManceraPina2020}, all originally identified in the ALFALFA survey by \citet{Leisman2017}. For AGC~114905, the updated outer circular speed ($v_{\rm obs}$) is adopted from \citet{ManceraPina2022}. The most recent resolved discussion of this galaxy is given by \citet{ManceraPina2024}, especially regarding the role of inclination and the full rotation-curve decomposition. They have baryonic masses $M_{\rm bar}$ near $10^9\,\mathrm{M_\odot}$, large gas fractions and extended discs. Quantitatively, their gas-to-stellar mass ratios span $M_{\rm gas}/M_\star \simeq 2.4$--$29.1$, their optical scale lengths range from $1.79$ to $4.15$ kpc, and their outer kinematic radii cover $r_{\rm sys}\simeq 8.0$--$10.8$ kpc. Previous analyses already showed that these galaxies are difficult to reconcile with standard MOND prescriptions, especially when their low outer circular speeds are compared with the baryonic mass inferred from the gas-rich discs \citep{ManceraPina2019,ManceraPina2020,ManceraPina2022}. The cold gas mass in these systems is usually estimated from the integrated 21-cm \HI\ line flux and then corrected for helium; in the underlying data compilation, the baryonic mass is dominated by the gas component. Table~\ref{tab:sample} summarizes the main observed properties of the sample, with central values and uncertainty intervals of $\log M_{\rm bar}$ and $v_{\rm obs}$ at the outermost measured radius $r_{\rm sys}$.

These six galaxies define the homogeneous likelihood sample used in this study. Although AGC~114905 has the best resolved rotation curve in the sample, it is treated here through the same outer-radius statistic as the other five galaxies in order to keep the comparison homogeneous. Resolved Apertif data provide additional examples of UDG-like systems with resolved kinematics \citep{Siljeg2024}, but these are kept as a separate exploratory extension rather than folded into the main six-galaxy likelihood.

\begin{table*}
\centering
\renewcommand{\arraystretch}{1.20}
\caption{Properties of the gas-rich UDG sample used in this study.}
\label{tab:sample}
\scriptsize
\begin{tabular}{lcccccccccc}
\toprule
Name\,$^{\rm a}$ &
Distance\,$^{\rm b}$ &
Inclination\,$^{\rm c}$ &
$r_d$\,$^{\rm d}$ &
$\log(M_{\mathrm{bar}}/M_{\odot})$ &
$\log(M_{\star}/M_{\odot})$ &
$M_{\rm gas}/M_{\star}$\,$^{\rm e}$ &
$v_{\mathrm{\rm circ}}$\,$^{\rm f}$ &
$r_{\rm sys}$\,$^{\rm g}$ &
$v_{\text{\tiny N},r_{\rm sys}}$\,$^{\rm h}$ \\
(AGC) & (Mpc) & (deg) & (kpc) &  &  &  & (km s$^{-1}$) & (kpc) & (km s$^{-1}$) \\
\midrule
114905 & 76 & 33 & $1.79 \pm 0.04$ & $9.21 \pm 0.19$ & $8.30 \pm 0.17$ & $7.1^{+2.3}_{-1.9}$ & $23^{+4}_{-6}$ & 8.02  & $29^{+7}_{-6}$ \\
122966 & 90 & 34 & $4.15 \pm 0.19$ & $9.21 \pm 0.13$ & $7.73 \pm 0.12$ & $29.1^{+7.0}_{-11.9}$ & $37^{+5}_{-6}$ & 10.80 & $25^{+4}_{-4}$ \\
219533 & 96 & 42 & $2.35 \pm 0.20$ & $9.36 \pm 0.21$ & $8.04 \pm 0.12$ & $19.7^{+8.8}_{-12.2}$ & $37^{+6}_{-5}$ & 9.78 & $32^{+9}_{-7}$ \\
248945 & 84 & 66 & $2.08 \pm 0.07$ & $9.05 \pm 0.19$ & $8.52 \pm 0.17$ & $2.4^{+0.8}_{-1.6}$ & $27^{+3}_{-3}$ & 8.55 & $24^{+6}_{-5}$ \\
334315 & 73 & 52 & $3.76 \pm 0.14$ & $9.25 \pm 0.16$ & $7.93 \pm 0.12$ & $19.7^{+5.9}_{-9.8}$ & $25^{+5}_{-5}$ & 8.49 & $30^{+6}_{-5}$ \\
749290 & 97 & 39 & $2.38 \pm 0.14$ & $9.17 \pm 0.15$ & $8.32 \pm 0.13$ & $6.1^{+1.7}_{-2.9}$ & $26^{+6}_{-6}$ & 8.47 & $27^{+5}_{-4}$ \\
\bottomrule
\end{tabular}

\vspace{0.6ex}
\parbox{\textwidth}{\footnotesize
$^{\rm a}$ Arecibo General Catalogue ID.\quad
$^{\rm b}$ Distance from \citet{Leisman2017}; the original compilation quotes an uncertainty of about $\pm 5$ Mpc.\quad
$^{\rm c}$ Inclination from \citet{ManceraPina2020}; the tabulated uncertainty is $\pm 5^\circ$.\quad
$^{\rm d}$ Optical disc scale length from \citet{ManceraPina2020}.\quad
$^{\rm e}$ Ratio between gas and stellar masses, with $M_{\rm gas}=1.33\,M_{\rm HI}$.\quad
$^{\rm f}$ Outer circular speed from the resolved \HI analyses of \citet{ManceraPina2020}, with AGC~114905 updated by \citet{ManceraPina2022}.\quad
$^{\rm g}$ Radius of the system defined here by the outermost ring of the published rotation curve.\quad
$^{\rm h}$ Published Newtonian disc-based circular speed inferred from the resolved baryonic distribution at $r_{\rm sys}$ \citep{ManceraPina2020}.}
\end{table*}

\section{Outer-radius test}

We tested the modelled circular speed at the radius $r_{\rm sys}$ within three frameworks. In the Newtonian case, we use the published disc-based outer-radius velocities $v_{\text{\tiny N},r_{\rm sys}}$ listed in Table~\ref{tab:sample}, derived from the resolved baryonic distribution rather than from a spherical approximation. For consistency, the same disc-based $v_{\text{\tiny N},r_{\rm sys}}$ is also used as the Newtonian input in Eqs.~\ref{eq:gammaeps} and \ref{eq:HMG}; the baryonic mass still enters the neighbourhood-density term $\epsilon_s^2(s)$ in Eq.~\ref{eq:epss}. The MOND speed with the MLS inverse interpolation function (Eq.~\ref{eq:MLS}) is \begin{equation}
v_\text{\tiny MOND}^2 = g_\text{\tiny MOND}\,r_{\rm sys}
= \hat{\nu}_{\rm MLS}(g_\text{\tiny N }/a_0)\,g_\text{\tiny N }\,r_{\rm sys},
\label{eq:vMOND}
\end{equation}
and, finally, the outer-radius circular velocity in HMG is written as
\begin{equation}
v_\text{\tiny HMG}^2 = g_\text{\tiny HMG}\,r_{\rm sys},
\label{eq:vHMG}
\end{equation}
where $g_\text{\tiny HMG}$ is the total centripetal acceleration of the HMG prescription in Eq.~\ref{eq:HMG}. For AGC~114905, where better resolved data exist, a full rotation-curve HMG fit remains a crucial next step and is discussed below.

For the likelihood scan, the observed velocities are compared with the HMG predictions at fixed $s$. The quantity minimised is the standard chi-square,
\begin{equation}
\begin{aligned}
\chi^2(s)&=\sum_i\left[\frac{v_{\text{\tiny HMG},i}(s)-v_{\text{\tiny obs},i}}{\sigma_{{\rm eff},i}}\right]^2,\\
\sigma_{{\rm eff},i}&=
\begin{cases}
\sigma_{+,i}, \;\; v_{\text{\tiny HMG},i}(s)\ge v_{\text{\tiny obs},i},\\
\sigma_{-,i}, \;\; v_{\text{\tiny HMG},i}(s)< v_{\text{\tiny obs},i},
\end{cases}
\end{aligned}
\label{eq:chisq}
\end{equation}
where $(\sigma_{+,i},\sigma_{-,i})$ are the published asymmetric velocity errors for galaxy $i$. Model-side uncertainties for the MOND and HMG comparisons were estimated through Monte Carlo propagation of the baryonic inputs in the accompanying reproducible scripts, while the Newtonian intervals follow the published resolved-disc values.

For the per-galaxy tension summary of Table~\ref{tab:velocities}, the comparison is expressed in terms of the combined uncertainty
\begin{equation}
\sigma_{{\rm comb},i} = \left(\sigma_{{\rm obs},i}^2+\sigma_{{\rm mod},i}^2\right)^{1/2},
\label{eq:sigmacomb}
\end{equation}
using the appropriate side of the asymmetric observational and model intervals according to the sign of $v_{\rm model}-v_{\rm obs}$.

\section{Results}

The HMG scan does not show an interior minimum for $s$. As expected, $\chi^2(s)$ decreases monotonically from very poor values at small $s$ towards an asymptote reached in the weak-coupling limit ($s \gg 1$). Numerically, the best achievable value is $\chi^2_{\min} \simeq 17.13$ for the fixed asymptotic branch applied to the six galaxies, with the minimum pushed to the upper edge of the scan range. A one-sided 95 per cent lower limit gives $s>43.9$, so the preferred solutions lie deeply in the asymptotic regime. In that limit, $\epsilon_s^2$ is already close to $1/6$ for all six galaxies and $v_\text{\tiny N } \gg v_\text{\tiny H}$ throughout the sample, effectively suppressing the direct cosmological contribution to the circular velocity. The scan is nevertheless informative because it shows that the data drive the model towards the asymptotic branch, $\epsilon_s^2 \approx 1/6$.

\begin{figure*}
\centering
\includegraphics[width=\textwidth]{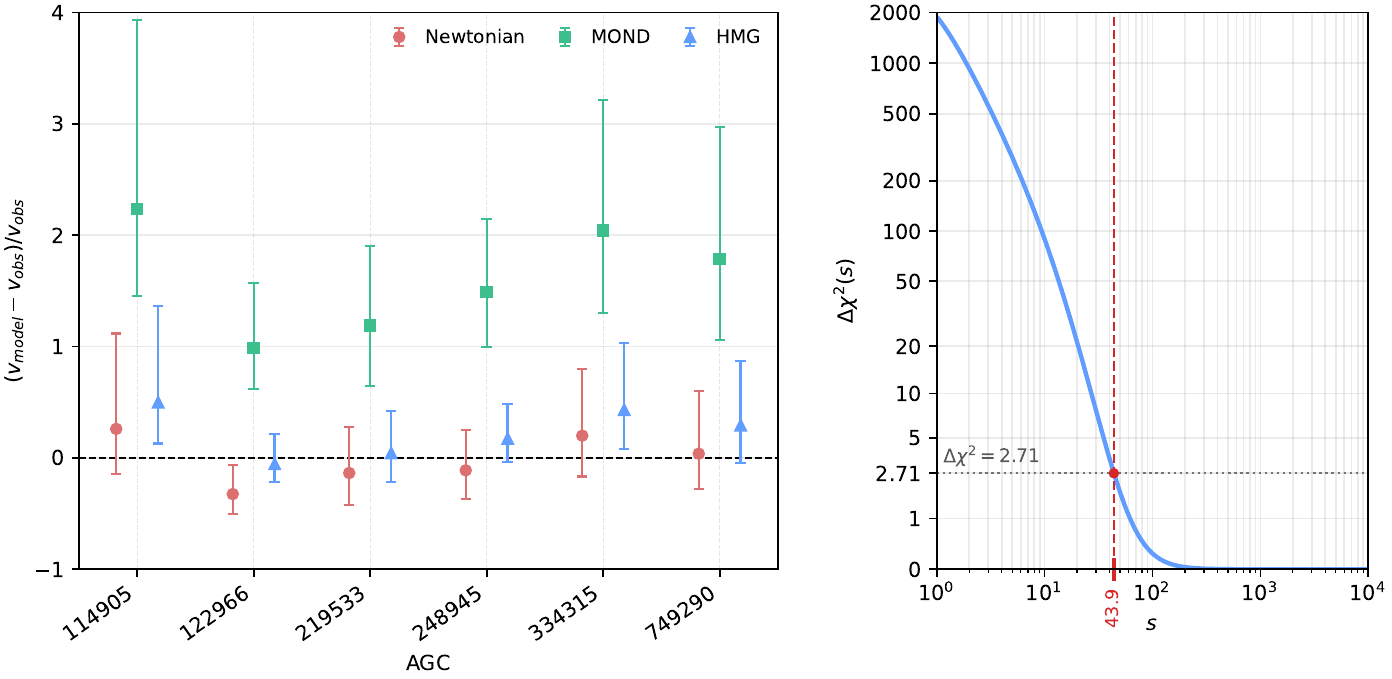}
\caption{Left: fractional velocity offsets with respect to the observed outer circular speed, including the propagated asymmetric intervals. The HMG asymptotic branch lies much closer to the data than the standard MLS MOND prediction, although the published Newtonian disc-based values remain the closest overall. Right: profile $\Delta\chi^2(s)$ for the HMG neighbourhood-scale factor, displayed on a $\ln(1+\Delta\chi^2)$ vertical scale. The curve decreases toward the asymptotic branch and gives the one-sided 95 per cent lower limit $s>43.9$.}
\label{fig:main_results}
\end{figure*}

The resulting hierarchy between models is clear. In the same outer-radius treatment, the published Newtonian disc-based comparison gives $\chi^2\simeq9.3$, slightly better than the present HMG prescription, while the MOND estimate is grossly inconsistent with the data ($\chi^2\simeq615.7$). HMG therefore does alleviate the strong MOND tension and brings the predicted velocities much closer to the Newtonian scale (Fig.~\ref{fig:main_results}).

\begin{table}
\centering
\caption{Observed outer circular speeds and model comparisons. The $z$ columns give the per-galaxy tension in units of the combined observational and model uncertainty.}
\label{tab:velocities}
\scriptsize
\renewcommand{\arraystretch}{1.18}
\setlength{\tabcolsep}{3.2pt}
\begin{tabular}{lccccccc}
\toprule
Galaxy & $v_{\rm obs}$ & $v_\text{\tiny N }$ & $z_\text{\tiny N }$ & $v_{\rm MOND}$ & $z_{\rm MOND}$ & $v_{\rm HMG,\infty}$ & $z_{\rm HMG,\infty}$ \\
 & (km s$^{-1}$) & (km s$^{-1}$) & ($\sigma$) & (km s$^{-1}$) & ($\sigma$) & (km s$^{-1}$) & ($\sigma$) \\
\midrule
114905 & $23^{+4}_{-6}$ & $29^{+7}_{-6}$ & 0.83 & $74.4^{+9.5}_{-8.2}$ & 5.66 & $34.5^{+5.7}_{-4.0}$ & 2.03 \\
122966 & $37^{+5}_{-6}$ & $25^{+4}_{-4}$ & -1.66 & $73.6^{+6.1}_{-5.7}$ & 4.83 & $35.0^{+2.6}_{-2.2}$ & -0.31 \\
219533 & $37^{+6}_{-5}$ & $32^{+9}_{-7}$ & -0.49 & $81.0^{+12.0}_{-10.2}$ & 3.72 & $38.6^{+7.0}_{-4.8}$ & 0.21 \\
248945 & $27^{+3}_{-3}$ & $24^{+6}_{-5}$ & -0.45 & $67.2^{+8.3}_{-7.3}$ & 5.09 & $31.7^{+4.0}_{-2.9}$ & 1.13 \\
334315 & $25^{+5}_{-5}$ & $30^{+6}_{-5}$ & 0.71 & $76.1^{+8.2}_{-7.0}$ & 5.91 & $35.8^{+4.8}_{-3.5}$ & 1.77 \\
749290 & $26^{+6}_{-6}$ & $27^{+5}_{-4}$ & 0.14 & $72.4^{+7.0}_{-6.4}$ & 5.27 & $33.6^{+3.8}_{-3.0}$ & 1.13 \\
\bottomrule
\end{tabular}
\end{table}

However, the asymptotic HMG prediction overshoots the observed outer speed by $\sim8$--$12\kms$ for AGC~114905, AGC~334315 and AGC~749290, while AGC~122966 and AGC~219533 remain comparatively well reproduced (Table~\ref{tab:velocities}). In terms of the combined model-versus-observation tension, HMG lies in the range $0.2$--$2.0\sigma$, very close to the Newtonian range of $0.1$--$1.7\sigma$, whereas MOND remains discrepant at $3.7$--$5.9\sigma$.

A useful complementary summary is obtained from the combined tensions of Table~\ref{tab:velocities}. Summing the squared $z$ values gives $\chi^2_{\rm comb}\simeq 4.4$ for Newton and $\chi^2_{\rm comb}\simeq 10.0$ for HMG, so that Newton remains favoured, but only at the level of a relative Gaussian weight of order 6 per cent for HMG. This is still a clear contrast with MOND, whose tensions remain much larger galaxy by galaxy.

In fractional terms, MOND overshoots all six galaxies by order-unity factors, whereas HMG reduces the offsets to the range $\sim -0.05$ to $0.50$ (Fig.~\ref{fig:main_results} left). The remaining HMG tension is driven mainly by AGC~114905, AGC~334315, and AGC~749290.

\section{Discussion}

\paragraph*{Limitations of the test.}

The main limitation of the present compact test is that it does not yet propagate the full distance and inclination uncertainties into the outer velocity likelihood. The reason is practical rather than conceptual: for several galaxies these uncertainties enter through the original 3D \HI modelling and are not published as homogeneous posterior samples for all six systems. A fully consistent treatment would therefore require re-modelling the underlying data cubes rather than simply inflating the one-point velocity errors. We state this limitation explicitly because the analyses of \citet{Lelli2024} and \citet{ManceraPina2024} show that distance and inclination can materially affect the interpretation of individual objects, particularly AGC~114905.

\paragraph*{Neighbourhood-scale branch.}

The scan is also restricted to the physically intended outer-neighbourhood domain $s\geq1$. If this restriction is artificially relaxed, the formal expression approaches a quasi-Newtonian regime at very small $s$: for example, at $s\simeq10^{-3}$ the six $v_{\rm HMG}$ values are essentially identical to the published disc-based Newtonian values, yielding $\chi^2\simeq9.2$. At $s=10^{-2}$ the velocities rise slightly to $30.0,26.5,33.2,25.3,31.1$, and $28.1\kms$, giving $\chi^2\simeq8.6$. This numerical behaviour is not interpreted here as a preferred HMG solution, because $s<1$ corresponds to evaluating the neighbourhood density inside $r_{\rm sys}$ rather than on an outer system scale. This behaviour follows directly from Eq.~\ref{eq:gammaeps}: writing $\xi=\sqrt{2}v_{\rm N}/(\epsilon_s v_{\rm H})$, the modulation contains $|(\xi^2-1)/(\xi^2+1)|$, so both $\xi\gg1$ and $\xi\ll1$ approach quasi-Newtonian tails around a maximum HMG enhancement near $\xi\simeq1$. Thus the artificial $s\to0$ limit, for which $\epsilon_s\gg1$, is not a new preferred branch but the opposite quasi-Newtonian tail. The physically adopted domain therefore remains $s\geq1$, for which the minimum is reached only on the asymptotic branch and gives $\chi^2\simeq17.1$.

\paragraph*{Resolved follow-up.}

The outer-radius statistic used here is designed to test the asymptotic behaviour of the HMG prescription in a homogeneous way across the six galaxies. This is useful precisely because gas-rich UDGs probe a low-acceleration, extended-disc regime where MOND and HMG differ most clearly. However, a single outer point cannot determine how the model connects the inner baryon-dominated region, the transition regime, and the outer asymptotic branch. AGC~114905, which has the best resolved rotation curve in the sample, is therefore the natural target for a full resolved HMG fit. Such an analysis would test not only the outer velocity scale considered here, but also the radial shape of the predicted curve and its connection with the broader dynamical behaviour of gas-rich galaxies.

\paragraph*{Exploratory Apertif systems.}

The additional resolved Apertif UDGs of \citet{Siljeg2024} can be used as an exploratory consistency check, but they are not folded into the main likelihood because published disc-based Newtonian outer velocities are not available in the same form as for the Mancera Pi\~na et al. sample. Instead, Table~\ref{tab:siljeg_extension} lists the spherical outer-radius estimate $v_{\rm N,sph}=(GM_{\rm bar}/R_{\rm out})^{1/2}$. This approximation is useful as a first diagnostic because it reproduces the original six-galaxy Newtonian entries to within a few per cent, but it is still labelled explicitly as spherical. The two robust UDGs with resolved outer speeds in \citet{Siljeg2024}, AHCJ1308+5437 and AHCJ2207+4008, do not extend the low-velocity tension: their observed velocities are higher than $v_{\rm N,sph}$ and also higher than the asymptotic HMG estimate. As an exploratory diagnostic, Table~\ref{tab:siljeg_extension} therefore keeps the asymptotic HMG prediction for comparison and adds the HMG velocity obtained when fitting a single common neighbourhood-scale factor to the two galaxies. In the allowed $s\geq1$ branch this gives $s\simeq4.84$, close to the individual values $s\simeq6.20$ and $s\simeq3.92$. Thus these two systems would favour a finite-neighbourhood solution rather than the asymptotic branch used in the main six-galaxy analysis. A third object, AHCJ2207+4143, has resolved kinematics but is not included in this minimal extension because the optical classification and photometric fit are less secure.

\begin{table}
\centering
\caption{Exploratory outer-radius estimates for two additional resolved Apertif UDGs from \citet{Siljeg2024}. These values are not included in the main likelihood. The final two columns compare the asymptotic HMG branch with the finite-neighbourhood solution obtained from a single common fitted value $s=4.84$.}
\label{tab:siljeg_extension}
\scriptsize
\setlength{\tabcolsep}{2.1pt}
\begin{tabular}{lcccccc}
\toprule
Galaxy & $\log M_{\rm bar}$ & $R_{\rm out}$ & $v_{\rm obs}$ & $v_{\rm N,sph}$ & $v_{\rm HMG,\infty}$ & $v_{\rm HMG}(s=4.84)$ \\
 &  & (kpc) & (km s$^{-1}$) & (km s$^{-1}$) & (km s$^{-1}$) & (km s$^{-1}$) \\
\midrule
AHCJ1308+5437 & 9.63 & 17.3 & $70.4^{+9.3}_{-5.4}$ & 32.6 & 47.0 & 77.1 \\
AHCJ2207+4008 & 9.72 & 17.1 & $87.9^{+24.1}_{-9.5}$ & 36.3 & 49.1 & 81.3 \\
\bottomrule
\end{tabular}
\end{table}

The neighbourhood scale inferred from these two exploratory systems is close to that found for ordinary galaxy rotation curves by \citet{Monjo_2025_distinct}, for which the typical-density radius is $r_{\rm typ}\sim4r_{\max}$, corresponding here to $s\sim4$ if $r_{\max}$ is identified with the outer system radius. The contrast is therefore informative: the Apertif UDGs favour finite-neighbourhood values $s\simeq3.9$--$6.2$, whereas the original six-galaxy sample requires $s>43.9$ and drives the current outer-radius prescription to its asymptotic weak-coupling branch.

\section{Conclusions}

Gas-rich UDGs are an appropriate target because they provide a clean and potentially high-impact falsification test. Applied to the six-galaxy sample of \citet{ManceraPina2020}, the current outer-radius HMG implementation reproduces the observed central values adequately for only two of the six galaxies. The scan pushes the model to the asymptotic weak-coupling branch, with a one-sided 95 per cent lower limit $s>43.9$, and that branch still predicts velocities that are relatively high for most galaxies. The present outer-radius HMG prescription is not yet favoured over the published Newtonian disc-based comparison, but it closes much of the gap between MOND and the data. In that sense, gas-rich UDGs already act as a useful discriminant: they strongly disfavour standard MOND, while separating only more moderately between Newton and HMG.

The main implication is constructive. Because of their diffuse mass distributions, HMG may accommodate gas-rich UDGs more successfully with at least one of three upgrades: i) a resolved disc calculation rather than a single outer-radius estimate, ii) a revised relation between $\gamma_{\rm sys}$ and the neighbourhood density $\rho_\text{nei}(s)$, or iii) a full hierarchical treatment of observational systematics. The additional Apertif systems discussed above point in the same direction: they are not part of the homogeneous likelihood, but they show that finite-neighbourhood solutions can differ substantially from the asymptotic branch. Until then, gas-rich UDGs should be regarded as a non-trivial challenge for the present HMG phenomenology and as a useful discriminant between HMG and MOND-like prescriptions. The present manuscript is therefore best framed as a compact comparative stress test of the current prescription, rather than as a definitive exclusion of the wider theory. A resolved HMG confrontation with the full AGC~114905 rotation curve is the most direct next step.

\section*{Data Availability}
The reproducible scripts used for this comparison are publicly available in the GitHub repository associated with this work, \url{https://github.com/robertmonjo/hmg-udg-repro}, with an archived software release available through Zenodo: \url{https://doi.org/10.5281/zenodo.19321452}.

\section*{Acknowledgements}
We thank the authors of the UDG studies for making the relevant galaxy properties publicly available.

\bibliographystyle{mnras}
\bibliography{references}

@article{Leisman2017,
  author = {Leisman, Lukas and Haynes, Martha P. and Janowiecki, Steven and Hallenbeck, Gregory and J{\'o}zsa, Gyula and Giovanelli, Riccardo and Adams, Elizabeth A. K. and Bernal Neira, David and Cannon, John M. and Janesh, William F. and Rhode, Katherine L. and Salzer, John J.},
  title = {(Almost) Dark Galaxies in the {ALFALFA} Survey: Isolated {H I}-bearing Ultra-diffuse Galaxies},
  journal = {Astrophysical Journal},
  year = {2017},
  volume = {842},
  pages = {133},
  doi = {10.3847/1538-4357/aa7575}
}

@article{ManceraPina2019,
  author = {Mancera Pi{\~n}a, Pavel E. and Fraternali, Filippo and Adams, Elizabeth A. K. and Marasco, Antonino and Oosterloo, Tom and Oman, Kyle A. and Leisman, Lukas and di Teodoro, Enrico M. and Posti, Lorenzo and Battipaglia, Michael and Cannon, John M. and Gault, Lexi and Haynes, Martha P. and Janowiecki, Steven and McAllan, Elizabeth and Pagel, Hannah J. and Reiter, Kameron and Rhode, Katherine L. and Salzer, John J. and Smith, Nicholas J.},
  title = {Off the Baryonic Tully--Fisher Relation: A Population of Baryon-dominated Ultra-diffuse Galaxies},
  journal = {Astrophysical Journal Letters},
  year = {2019},
  volume = {883},
  number = {2},
  pages = {L33},
  doi = {10.3847/2041-8213/ab40c7}
}

@article{ManceraPina2020,
  author = {Mancera Pi{\~n}a, Pavel E. and Fraternali, Filippo and Oman, Kyle A. and Adams, Elizabeth A. K. and Bacchini, Cecilia and Marasco, Antonino and Oosterloo, Tom and Pezzulli, Gabriele and Posti, Lorenzo and Leisman, Lukas and Cannon, John M. and di Teodoro, Enrico M. and Gault, Lexi and Haynes, Martha P. and Reiter, Kameron and Rhode, Katherine L. and Salzer, John J. and Smith, Nicholas J.},
  title = {Robust {H I} Kinematics of Gas-rich Ultra-diffuse Galaxies: Hints of a Weak-feedback Formation Scenario},
  journal = {Monthly Notices of the Royal Astronomical Society},
  year = {2020},
  volume = {495},
  number = {4},
  pages = {3636--3655},
  doi = {10.1093/mnras/staa1256}
}

@article{ManceraPina2022,
  author = {Mancera Pi{\~n}a, Pavel E. and Fraternali, Filippo and Oosterloo, Tom and Adams, Elizabeth A. K. and Oman, Kyle A. and Leisman, Lukas and Bacchini, Cecilia and Marasco, Antonino and Campo, Daniela and Gault, Lexi and Haynes, Martha P. and Reiter, Kameron and Rhode, Katherine L. and Salzer, John J.},
  title = {No Need for Dark Matter: Resolved Kinematics of the Ultra-diffuse Galaxy {AGC} 114905},
  journal = {Monthly Notices of the Royal Astronomical Society},
  year = {2022},
  volume = {512},
  number = {3},
  pages = {3230--3242},
  doi = {10.1093/mnras/stab3491}
}

@article{Lelli2024,
  author = {Lelli, Federico},
  title = {Gas-rich ``ultra-diffuse'' galaxies are consistent with the baryonic Tully--Fisher relation and with Milgromian dynamics},
  journal = {Astronomy \& Astrophysics},
  year = {2024},
  volume = {689},
  pages = {L3},
  doi = {10.1051/0004-6361/202451289}
}

@article{Siljeg2024,
  author = {\v{S}iljeg, B. and Adams, E. A. K. and Fraternali, F. and Hess, K. M. and Oosterloo, T. A. and Marasco, A. and Adebahr, B. and D{\'e}nes, H. and Garrido, J. and Lucero, D. M. and Mancera Pi{\~n}a, P. E. and Moss, V. A. and Parra-Roy{\'o}n, M. and Ponomareva, A. A. and S{\'a}nchez-Exp{\'o}sito, S. and van der Hulst, J. M.},
  title = {Photometry and kinematics of dwarf galaxies from the Apertif {H I} survey},
  journal = {Astronomy \& Astrophysics},
  year = {2024},
  volume = {692},
  pages = {A217},
  doi = {10.1051/0004-6361/202449923}
}

@article{ManceraPina2024,
  author = {Mancera Pi{\~n}a, Pavel E. and Golini, Giulia and Trujillo, Ignacio and Montes, Mireia},
  title = {Exploring the nature of dark matter with the extreme galaxy {AGC} 114905},
  journal = {Astronomy \& Astrophysics},
  year = {2024},
  volume = {689},
  pages = {A344},
  doi = {10.1051/0004-6361/202450230}
}

@article{Milgrom1983,
  author = {Milgrom, Mordehai},
  title = {A Modification of the Newtonian Dynamics as a Possible Alternative to the Hidden Mass Hypothesis},
  journal = {Astrophysical Journal},
  year = {1983},
  volume = {270},
  pages = {365--370}
}

@article{McGaugh2016,
  author = {McGaugh, Stacy S. and Lelli, Federico and Schombert, James M.},
  title = {Radial Acceleration Relation in Rotationally Supported Galaxies},
  journal = {Physical Review Letters},
  year = {2016},
  volume = {117},
  pages = {201101},
  doi = {10.1103/PhysRevLett.117.201101}
}

@article{BanikZhao2022,
  author = {Banik, Indranil and Zhao, Hongsheng},
  title = {From Galactic Bars to the Hubble Tension: Weighing up the Astrophysical Evidence for Milgromian Gravity},
  journal = {Symmetry},
  year = {2022},
  volume = {14},
  number = {7},
  pages = {1331},
  doi = {10.3390/sym14071331}
}

@article{Monjo2023CQG,
  author = {Monjo, Robert},
  title = {Galaxy Rotation Curve in Hyperconical Universes: A Natural Relativistic {MOND}},
  journal = {Classical and Quantum Gravity},
  year = {2023},
  volume = {40},
  number = {21},
  pages = {215003},
  doi = {10.1088/1361-6382/ad0422}
}

@article{Monjo_2025_distinct,
  author = {Monjo, Robert and Banik, Indranil},
  title = {Distinct Acceleration Relations of Galaxies and Galaxy Clusters from Hyperconical Modified Gravity},
  journal = {Astrophysical Journal},
  year = {2025},
  volume = {992},
  number = {1},
  pages = {35},
  doi = {10.3847/1538-4357/adfcc0}
}

@article{Monjo_2025_weak,
  author = {Monjo, Robert},
  title = {Comparison of {HMG} and Flat Rotation Velocities Inferred from Galaxy--Galaxy Weak Lensing},
  journal = {Astrophysical Journal},
  year = {2025},
  volume = {982},
  number = {2},
  pages = {70},
  doi = {10.3847/1538-4357/adb8d7}
}

@article{Monjo_2025hydro,
  author = {Monjo, Robert},
  title = {Hydrostatic Equilibrium Constraints in X-COP Galaxy Clusters: Testing Hyperconical Modified Gravity in Hot-gas-dominated Regions},
  journal = {Astrophysical Journal},
  year = {2025},
  volume = {981},
  number = {2},
  pages = {195},
  doi = {10.3847/1538-4357/adb723}
}

@article{vanPutten2015,
  author = {{van Putten}, Maurice H. P. M.},
  title = {A Holographic Bound on the Total Number of Computations in the Visible Universe},
  journal = {International Journal of Modern Physics D},
  year = {2015},
  volume = {24},
  number = {3},
  pages = {1550024},
  doi = {10.1142/S0218271815500248}
}

@article{vanPutten2024,
  author = {{van Putten}, Maurice H. P. M.},
  title = {Galaxy Dynamics Tracing Quantum Cosmology beyond {$\Lambda$CDM} below the de {Sitter} Scale of Acceleration},
  journal = {Chinese Journal of Physics},
  year = {2024},
  volume = {91},
  pages = {377--381},
  doi = {10.1016/j.cjph.2024.07.040}
}

\bsp
\label{lastpage}
\end{document}